\documentclass[graybox]{svmult}

\usepackage{amsmath,amsfonts,amssymb}
\usepackage{mathptmx}       
\usepackage{helvet}         
\usepackage{courier}        
\usepackage{type1cm}        
\usepackage{makeidx}         
\usepackage{graphicx}        
\usepackage{multicol}        
\usepackage[bottom]{footmisc}

\usepackage{aas_macros}

\makeindex             

\begin{document}

\title*{Supermassive Black Hole Binaries: \\The Search Continues}
\author{Tamara Bogdanovi\'c}
\institute{Tamara Bogdanovi\'c \at Center for Relativistic Astrophysics, School of Physics, Georgia Institute of Technology, Atlanta, GA 30332-0430, \email{tamarab@gatech.edu}}
%
%
\maketitle

\abstract*{Gravitationally bound supermassive black hole binaries (SBHBs) are thought to be a natural product of galactic mergers and growth of the large scale structure in the universe. They however remain observationally elusive, thus raising a question about characteristic observational signatures associated with these systems. In this conference proceeding I discuss current theoretical understanding and latest advances and prospects in observational searches for SBHBs.}

\abstract{Gravitationally bound supermassive black hole binaries (SBHBs) are thought to be a natural product of galactic mergers and growth of the large scale structure in the universe. They however remain observationally elusive, thus raising a question about characteristic observational signatures associated with these systems. In this conference proceeding I discuss current theoretical understanding and latest advances and prospects in observational searches for SBHBs.}

\section{Theoretical background: formation and evolution of SBHBs}
\label{sec:1}

Over the past two decades it became apparent that the evolution of galaxies and their supermassive black holes (SBHs) is intricately connected \cite{ferrarese00,gebhardt00,tremaine02}. It has also been known for a while that galaxies evolve through mergers \cite{white78,schw86,bh92, bh96, volonteri03,dimatteo08a}, thus raising a question: {\it when galaxies merge, what happens to their SBHs?}  In the aftermath of a generic merger the two parent galaxies form a new stellar bulge and their SBHs find themselves at a mutual separation of $\sim$kpc. At this stage, widely separated SBHs are gravitationally bound to the surrounding gas and stars but not to one another. In this proceeding I will refer to the dual SBHs in this phase of evolution as {\it pairs} and to the gravitationally bound SBHBs as {\it binaries}. 

Theoretical studies have established that evolution of SBH pairs from $\sim$kpc to smaller scales is determined by gravitational interactions of individual black holes with their environment (see the groundbreaking work by \cite{bbr80} for description and \cite{mayer13} for a most recent review). These include interaction of the SBHs with their own wakes of stars and gas, also known as the dynamical friction \cite{chandra43, ostriker99,mm01,escala04}, as well as the scattering of the SBHs by massive gas clouds and spiral arms produced by the local and global dynamical instabilities during the merger \cite{fiacconi13}. During these interactions the SBHs exchange orbital energy and angular momentum with the ambient medium and can in principle grow though accretion  \cite{escala04, escala05, dotti06, dotti07, dotti09, callegari11, khan12, chapon13}. These factors determine the SBH dynamics and whether they evolve to smaller separations to form a gravitationally bound binary. For example, \cite{callegari09,callegari11} find that SBH pairs with mass ratios $q<0.1$ are unlikely to form binaries within a Hubble time at any redshift. On the other hand SBH pairs with initially unequal masses can evolve to be more equal-mass, through preferential accretion onto a smaller SBH. It is therefore likely that SBH pairs with $q\gtrsim0.1$ form a parent population of bound binaries at smaller separations.

Gravitationally bound binary forms at the point when the amount of gas and stars enclosed within its orbit becomes comparable to the mass of the SBHB.  For a wide range of host properties and SBH masses this happens at orbital separations $\lesssim 10$~pc \cite{mayer07,dotti07,khan12}. The subsequent rate of binary orbital evolution sensitively depends on the nature of gravitational interactions that it experiences and is still an area of active research often abbreviated as  {\it the last parsec problem}.  The name refers to a possible slow-down and stalling in the orbital evolution of the parsec-scale SBHBs driven by the inefficient interactions with stars \cite{mm01} and gas \cite{escala05}. If present, a consequence of this effect would be that a significant fraction of SBHBs in the universe should reside at orbital separations of $\sim 1$pc. Several recent theoretical studies that focus on the evolution of {\it binaries in predominantly stellar backgrounds} however report that evolution of binaries to much smaller scales continues unhindered \cite{berczik06, preto11, khan11,khan12a,khan13}, although the agreement about the leading physical mechanism responsible for the evolution is still not universal \cite{vasiliev14}. 

{\it SBH binaries in predominantly gaseous environments} have also been a topic of a number of theoretical studies \cite{an05, macfadyen08, cuadra09, hayasaki09,roedig12,shi12,noble12,kocsis12a, kocsis12b,dorazio13,farris14}. They find that binary torques can truncate the sufficiently cold circumbinary disks and create an inner low density cavity by evacuating the gas from the central portion of the disk (see \cite{lp79} and references above). As the binary orbit decays, the inner rim of the disk follows it inward until the timescale for orbital decay by gravitational radiation becomes shorter than the viscous timescale\footnote{The time scale on which the angular momentum is transported outwards through the disk.} of the disk \cite{an05}. At that point, the rapid loss of orbital energy and angular momentum through gravitational radiation cause the binary to detach from the circumbinary disk and to accelerate towards coalescence. This final phase of binary evolution has been captured in a series of investigations based on fully relativistic particle \cite{vanmeter10}, electrodynamic and (magneto)hydrodynamic  simulations of coalescing binaries \cite{palenzuela09, bode10,farris10, palenzuela10a, palenzuela10,moesta10, moesta12, bode12,farris11,alic12,giaco12,gold14}.

Through its dependance on the viscous time scale, the orbital evolution of a gravitationally bound SBHB in the circumbinary disk sensitively depends on the thermodynamic properties of the disk. These are uncertain, as they are still prohibitively computationally expensive to model from the first principles and are unconstrained by observations. More specifically, the thermodynamics of the disk is determined by the binary dynamics but also by the presence of magnetic field and radiative heating and cooling. While the role of magnetic field is beginning to be explored in some simulations, a fully consistent calculation of radiative heating and cooling is still beyond the computational reach.   

Regardless of whether gravitationally bound SBH binaries evolve in mostly stellar or gas rich environments, the exchange of angular momentum with the ambient medium is likely to result in eccentric SBHB orbits \cite{an05,cuadra09,roedig11,sesana11,chapon13}. An interesting implication of this finding is that eccentric binaries that evolve all the way to the gravitational wave (GW) coalescence may leave a clear imprint of the orbital eccentricity in their emitted waveforms, detectable by the future space-based GW observatories, such as eLISA \cite{key11,as13}. Another property of astrophysical SBHs that will be possible to measure to high precision and high redshifts from eLISA observations are the magnitudes and orientations of the SBH spins prior to the coalescence \cite{as13}. This is an exciting prospect as these would complement the existing spin estimates for about two dozen single SBHs, based on measurements of the relativistically broadened X-ray FeK$\alpha$ emission line profiles (see \cite{reynolds13} for a recent review of this method).  

Our understanding of spin magnitudes and orientations in {\it binary} SBHs on the other hand relies mostly on theoretical considerations. Interest in this topic was triggered by the prediction of numerical relativity that coalescence of SBHs with certain spin configurations can lead to the ejection of a newly formed SBH from its host galaxy. This effect arises due to the asymmetry in emission of GWs in the final stages of a SBH merger and can lead to a GW kick of up to $\sim5000\,{\rm km\,s^{-1}}$ \cite{campanelli07,lousto11}. In majority of the binary configurations however, the GW kick velocity was found to be lower and is minimized whenever the SBH spin axes are aligned with the binary orbital axis.

Several subsequent theoretical studies found that accretion and gravitational torques can act to align the spin axes of SBHs evolving in gas reach environments and in such way minimize the GW recoil as long as the SBHs are orbiting within a rotationally supported, moderately geometrically thik disk \cite{bogdanovic07,dotti10,dotti13,sorathia13,miller13} (see however \cite{lodato13} for a different view).  The mutual SBH spin alignment is on the other hand not expected to happen in gas poor environments, geometrically thick, turbulent and magnetically dominated disks \cite{fragile05,fragile07,mckinney13} hence, allowing a possibility that runaway SBHs and empty nest galaxies may exist. This realization stimulated lots of research activity on the astrophysical implications of the GW kick \cite{bonning07,volonteri07,schnittman07,sesana07,loeb07,gualandris08,komossa08a,shields08,korn08,berti08,schnittman08,hb08,komossa08,blecha08,volonteri08,oneill09,devecchi09,merritt09,volonteri10,robinson10,lovelace10,corrales10,guedes11,blecha11,nixon11,nixon12,lousto12,ponce12,komossa12,stone12,shields13,king13,gerosa14}.  I will not dwell  further on recoiling SBHs in this proceeding except to note that some signatures of recoiling SBHs can coincide with those of the subarsec scale SBHBs (see Sections~\ref{sec:2.1} and \ref{sec:2.3}), effectively allowing to accomplish two searches with one observational strategy.

\section{Observational evidence for SBHBs}
\label{sec:2}

The key characteristic of gravitationally bound SBHBs is that they are observationally elusive and expected to be intrinsically rare. While the frequency of binaries is uncertain and dependent on their unknown rate of evolution on small scales (see previous section), theorists estimate that a fraction $<10^{-3}$ of active galactic nuclei (AGNs) at redshift $z<0.7$ may host SBHBs \cite{volonteri09}. This result has two important implications: (a) any observational search for SBHBs must involve a large sample of AGNs, making the archival data from large surveys of AGNs an attractive starting point and (b) observational technique used in a search must be able to distinguish signatures of binaries from those of AGNs powered by single SBHs. In the following sections I describe the application of imaging, photometric, and spectroscopic techniques, starting with the more direct methods and progressing towards less direct. I focus on the techniques that have been commonly used in SBHB searches and direct the reader to \cite{dotti12,schnittman13} for recent reviews of a broad range of SBHB signatures proposed in the literature.

\subsection{Direct imaging of double nuclei}
\label{sec:2.1}
\begin{figure}[b]
\sidecaption
\includegraphics[trim=10 70 30 35, clip, scale=.3,]{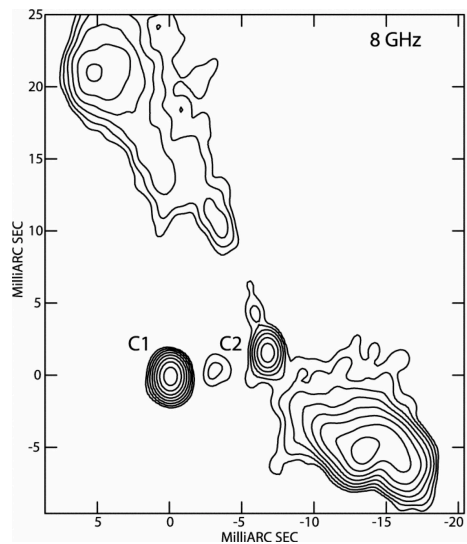}
\caption{VLBA image of the SBHB candidate in the radio galaxy 0402+379 observed at the frequency of 8GHz. Two compact radio cores, C1 and C2, are separated by 7.3~pc on the sky and are thought to harbor SBHs. Figure adapted from \cite{rodriguez06}.}
\label{fig1}      
\end{figure}

A clearest manifestation of a parsec scale SBHB is an image of a binary AGN which forms a gravitationally bound system (as opposed to an accidental projection on the sky). A practical obstacle in the detection of such objects arises from their small angular separation on the sky: for example, a parsec-scale binary at a moderate redshift of $z\approx0.2$ subtends an angle of only $\sim3$~mas on the sky (neglecting the projection effects). Such scales are below the angular resolution of most astronomical instruments, except the very long baseline interferometers (VLBI) used at radio wavelengths. 

A most convincing candidate for a SBHB in this class of objects was discovered in the radio galaxy 0402+379 by the Very Long Baseline Array (VLBA; Figure~\ref{fig1}) \cite{maness04,rodriguez06,rodriguez09,morganti09}. This object shows two compact radio cores at the projected separation of $7.3$~pc on the sky. Both cores are characterized by the flat radio spectra and have been identified as possible AGN based on this signature \cite{maness04,rodriguez06}.  While the SBHB candidate in the galaxy 0402+379 was discovered serendipitously, it demonstrated the power of radio interferometry in imaging of the small separation SBHs.

In a subsequent investigation, Burke et al. \cite{burke11} searched for binaries in the archival VLBI data. The search targeted spatially resolved, double radio-emitting nuclei with a wide range of orbital separations ($\sim3$pc$-5$kpc) among 3114 radio-luminous AGNs in the redshift range $0 < z \leq 4.715$. Another investigation in addition to SBHBs searched for the recoiling SBHs spatially offset from the centers of their host galaxies \cite{condon11}. The latter study is based on the VLBA 8GHz observations of 834 nearby radio-luminous AGN\footnote{Radio sources brighter than 100~mJy were selected based on the NVSS catalogue at 1.4GHz.} with typical distances of $\sim200$~Mpc. Neither search unearthed new instances of the double-radio nuclei, leaving several possible interpretations: (a) there is a true paucity of SBHBs, (b) SBHBs may be present but have low radio brightness, or (c) only one component of the binary is radio-bright and it may or may not show a detectable spatial offset relative to the center of the host galaxy. This points to difficulties in using the radio imaging as the primary technique to select the gravitationally bound SBHB candidates or their progenitors given their unknown radio properties. See however \cite{burke14} for a discussion of the optimal design of radio searches targeting the dual and binary SBHs.

\subsection{Photometric measurements of quasi-periodic variability}
\label{sec:2.2}

\begin{figure}[t]
\center{
\includegraphics[trim=120 270 130 270, clip, scale=0.8]{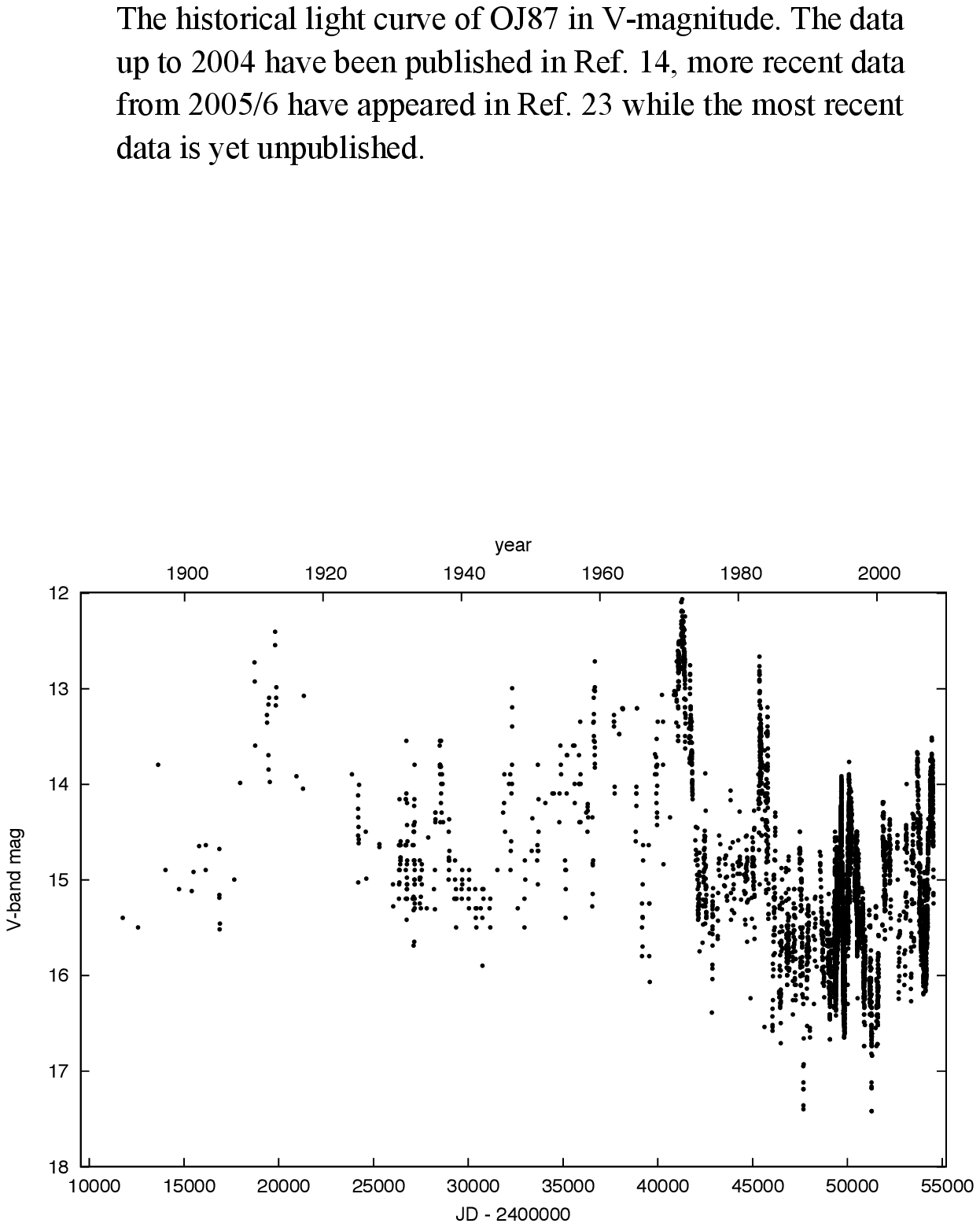}}
\caption{Historical light curve of OJ~287 in V-magnitude recorded over more than 100 years. Quasi-periodic outbursts with a period of $\sim12$~yr in the light curve of this object have been interpreted as a signature of the SBHB orbital motion. Figure from \cite{valtonen08}.}
\label{fig2}      
\end{figure}

The second most convincing line of evidence for the presence of a SBHB is a sustained periodic or nearly periodic variability on a time scale associated with the orbital period of the binary. This technique favors binaries with relatively short orbital periods, $P \lesssim 10$~yr, for which multiple cycles can be recorded in observations (for e.g., \cite{fan98,rieger00, depaolis02,liu14}). A well known example of a SBHB candidate in this category is a blazar OJ~287 which exhibits outburst activity in its optical light curve with a period close to 12 years (see Figure~\ref{fig2}), interpreted as a signature of the orbital motion \cite{valtonen08}. 

It is worth noting however that OJ~287 is unique among photometrically selected binary candidates because the first recorded data points in its light curve extend into the 19th century. Along similar lines, OJ~287 received an unprecedented level of observational coverage in modern times (from 1970s onwards), yielding a light curve with high frequency sampling. While indications of quasi-periodicity have been claimed in a handful of other objects, they are generally less pronounced and recorded over much shorter time span than in the case of OJ~287, thus preventing a strong case for SBHBs from being made in these sources. This observationally intensive technique may nevertheless play a more important role in the near future, as a number of high-cadence synoptic sky surveys come online. 

\subsection{Spectroscopic measurements of offset broad emission-lines}
\label{sec:2.3}

The least direct of the three methods relies on the spectroscopic detection of the Doppler-shift that arrises as a consequence of the binary orbital motion. It is based on a well established technique for detection of single- and double-line spectroscopic binary stars. In the double-line systems each offset line corresponds to one member of the binary, whereas in the single-line systems only one member is visible. In both classes of spectroscopic binaries, the lines are expected to oscillate about their rest frame wavelength on the time scale corresponding to the orbital period.

SBHs in the circumbinary disk phase (described in Section~\ref{sec:1}) can accrete by capturing gas from the inner rim of this disk. In the context of this model, the spectral lines are assumed to be associated with the gas accretion disks that are gravitationally bound to the individual SBHs \cite{bogdanovic08}. Given the velocities of the bound gas, the emission line profiles from the SBH mini-disks are expected to be Doppler-broadened, similar to the emission lines originating in the broad line regions (BLRs) of AGNs. Moreover, several theoretical studies have shown that for binary mass ratios $q < 1$ accretion occurs preferentially onto the lower mass object \cite{al96,gr00,hayasaki07}, rendering it potentially more luminous than the primary and indicating that some fraction of the SBHBs may appear as the single-line spectroscopic binaries. 

This realization motivated several searches for SBHBs based on the criterion that the culprit sources exhibit broad optical lines offset with respect to the rest frame of the host galaxy \cite{bogdanovic09a,dotti09a,bl09,tang09,decarli10, barrows11,tsal11,tsai13}. Because this effect is also expected to arise in the case of a recoiling SBH receding from its host galaxy, the same approach has been used to ÔÔflagÓ candidates of that type \cite{komossa08a,shields09,civano10, robinson10,lusso14}. The key advantage of the method is its simplicity, as spectra that exhibit Doppler shift signatures are relatively straightforward to select from large archival data sets, such as the Sloan Digital Sky Survey (SDSS). Its main complication however is that the Doppler shift signature is not unique to these two physical scenarios and complementary observations are needed in order to determine the true nature of the observed candidates \cite{bogdanovic09,popovic12}.

\begin{figure}[t]
\center{
\includegraphics[trim=120 190 160 240, clip, scale=0.75,angle=270]{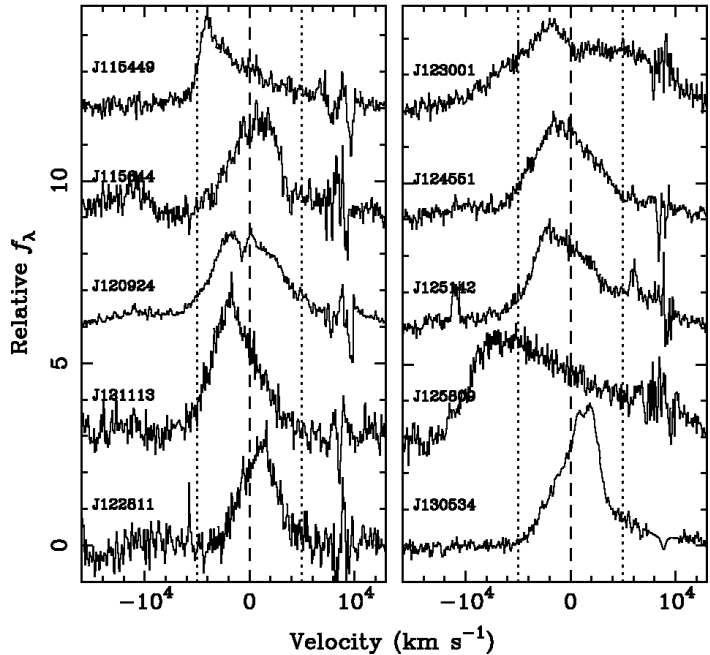}}
\caption{Broad H$\beta$ emission line profiles of selected SBHB candidates. Peakes of the asymmetric profiles are offset from the rest frame of the host galaxy (marked by the dashed line) by $\sim{\rm few}\times 10^3\;{\rm km\,s^{-1}}$. In the context of the SBHB model the broad profile is attributed to the emission from the accretion disk of the smaller SBH, and the offset is attributed to its orbital motion. Figure adapted from \cite{eracleous12}.}
\label{fig3}      
\end{figure}

To address this ambiguity a new generation of spectroscopic searches has been designed to monitor the offset of the broad emission line profiles over multiple epochs and target sources in which modulations in the offset are consistent with the binary motion \cite{eracleous12,bon12,decarli13,liu13,shen13,ju13}. For example, Eracleous et al. \cite{eracleous12} searched for $z<0.7$ SDSS quasars whose broad $H\beta$ lines are offset by $\gtrsim 1000\;{\rm km\,s^{-1}}$. Using this criterion they selected 88 quasars for observational followup from the initial catalog of $\sim 16,000$ objects. After the second and third epoch of observations of this sample, statistically significant changes in the velocity offset were found in 14 \cite{eracleous12} and 9 objects \cite{mathes14}, respectively, in broad agreement with theoretical predictions \cite{volonteri09}.  

Figure~\ref{fig3} shows several representative broad $H\beta$ profiles selected in this search. The profiles are asymmetric and have peaks offset by $\sim{\rm few}\times 10^3\;{\rm km\,s^{-1}}$ from the rest frame of the galaxy, as inferred from from the narrow emission lines (the narrow emission line components were subsequently removed from these profiles for clarity). If  velocity offset of $1000\;{\rm km\,s^{-1}}$ is interpreted as a signature of the orbital motion of a binary with mass $M\sim 10^8\,M_{\odot}$, it follows that the targeted population of SBHBs has the average orbital separation of $\sim 0.1$~pc and orbital period of $\sim 300$~yr. Given a long average orbital period and observational campaigns which typically span a time line about 10 years, it follows that searches of this type are in principle capable of monitoring a SBHB during a fraction of its orbital cycle but are in general not expected to record multiple cycles. 

A detection of an incomplete orbital cycle still leaves a possibility that other astrophysical processes may masquerade as binaries. For example, an orbiting hotspot produced by a local instability in the BLR of a galaxy, or outflows associated with the accretion disk may produce similar signatures and cannot be excluded. Consequently, the spectroscopic searches for SBHBs still require validation by another complementary observational technique.  Thanks to their efficiency in selection of the SBHB candidate samples, spectroscopic searches can be combined with the direct imaging of the binary nuclei (discussed in Section~\ref{sec:2.1}) and imaging of the host galaxy, in order to search for any signs of interaction. This two-step approach can in principle also be used to distinguish SBHBs from spatially and kinematically offset recoiling SBHs.

\section{Future theoretical and observational prospects}
\label{sec:3}

While selection of a well defined sample of SBHBs remains a principal goal in this research field, an equally important consideration is {\it what can be learned once such sample is available}. For example, the ongoing searches are based on the monitoring of the broad optical emission lines, which are expected to encode some information about the kinematics of the binary BLRs. It is thus plausible, although it remains to be demonstrated, that by analysis of these line profiles one can learn about the structure and thermodynamics of the circumbinary accretion flow, the very questions that are hampering the progress of theoretical models. More generally, a comparison with the spectroscopic data can provide a test of the underlying SBHB-in-a-circumbinary disk model as well as a constraint on the time scale for evolution of the gravitationally bound SBHBs. 

There are also other, fully developed observational techniques that can in principle be utilized to study SBHBs, once such a sample is defined. For example, {\it velocity-resolved reverberation mapping} is a method that has been successfully used to study the structure of the BLRs in several "conventional", single black hole AGNs \cite{bentz08,bentz09,bentz10,bentz10a,pancoast11,barth11,dietrich12,pancoast13}. 
The method relies on the measurement of the propagation of a light echo from the central source of the continuum radiation across the BLR. This is achieved through monitoring of the response of the permitted broad optical emission lines to the variations in the continuum. The approach yields constrains on the structure of orbits occupied by the emitting gas (i.e., circular or eccentric, inflowing or outflowing), as well as the SBH mass and inclination of the BLR relative to the observer's line of sight. Because it is observationally intensive, this technique can initially be applied to a subset of selected SBHB candidates with the goal to study the structure of their BLR regions. A typical monitoring campaign over several months of time would allow to capture the kinematics of the gas and factor out the variability due to the orbital motion of the binary, which for the parsec scale binaries occurs on much longer time scales (see Section~\ref{sec:2.3}).

Over the past two years a detection of time- and energy-resolved reverberation lags also became possible in the X-ray band and specifically, in the FeK$\alpha$ emission line region \cite{zoghbi12,kara13a,kara13b,kara14,cackett14,uttley14}. While key ideas are similar to the method used in the optical band, the main difference is that the FeK$\alpha$ line emission originates from the BLR very close to the SBH, within $\sim10\,r_g$\footnote{ $r_g = GM/c^2$ is the gravitational radius.} \cite{fabian89}.  The relativistically broadened profiles probe the structure of the innermost accretion disk region as well as the the spin of the SBH. In the context of SBHB model, this technique can in principle be applied to very close binaries at orbital separations of  $\sim 30-10^3\,r_g$, assuming that they can maintain the bright, X-ray emitting accretion disks within their orbit \cite{sesana12}. 
\begin{figure}[t]
\sidecaption[t]
\includegraphics[trim=190 310 190 310, clip, scale=0.92]{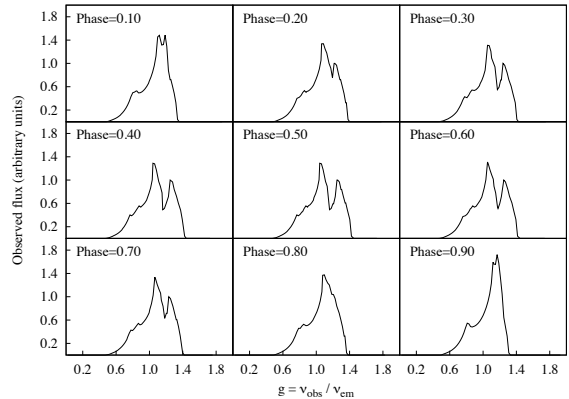}
\caption{Composite FeK$\alpha$ emission line profiles calculated for $q=1$ eccentric binary with orbital separation of $0.01$pc and accretion disks of equal luminosity. Profiles are shown as a function of the binary orbital phase in arbitrary flux units against frequency normalized to the rest frame frequency. Figure adapted from \cite{jovanovic13}.}
\label{fig4}      
\end{figure}

In reality, the X-ray emission properties of such close SBHB systems are virtually unknown beyond several theoretical models that predict the shape of the composite FeK$\alpha$ emission lines from binary BLRs \cite{mckernan13,jovanovic13}. Figure~\ref{fig4} shows a sequence of the FeK$\alpha$ line profiles as a function of the binary orbital phase calculated for an equal-mass binary with two disks of equal luminosity. This and other models illustrate that if the FeK$\alpha$ emission from SBHB systems can be observed, that line profiles will be distinct from those observed in the single SBH systems. Their detection however may be challenging due to the fact that these SBHB are relatively short lived and are expected to merge due to the emission of GWs on time scales of $10^4-10^6$~yr \cite{sesana12}.

Another promising channel for detection of this population of close SBHBs is with the pulsar timing arrays (PTAs) \cite{sesana08,shannon13,sesana13,burke13,arzu14}. This observational technique is based on the monitoring of a network of stable millisecond pulsars with the goal to measure a characteristic departure in the time of arrival of their radio pulses caused by GWs in the frequency range $(10^{-9}-10^{-7}{\rm Hz})$. While the PTA signal is expected to be dominated by the stochastic background of GWs emitted by a population of SBHBs with mass $>10^8\,M_{\odot}$ and $z<1$, detection of individual, particularly bright GW sources cannot be excluded \cite{babak12,ravi12,tanaka13,petiteau13}. A detection of the PTA SBHBs can constrain the dominant evolutionary scenario and possibly the accretion history of the most massive black holes in the universe. It conjunction with observations of the FeK$\alpha$ line profiles, this approach could provide the only 
opportunity to study the GW emitting SBHBs before the GW space-based observatories come online.

\section{Conclusions}
\label{sec:4}
 
Past ten years have marked a period of very active research on SBH pairs and binaries. The interest in them has been driven by a realization that SBHs play an important role in evolution of their host galaxies and also, by intention to understand the parent population of merging binaries because these are the prime targets of the long anticipated space-based GW observatories. While the future space-based GW detectors will undoubtedly transform our understanding of SBHBs, the electromagnetic signatures remain the only way to learn about this population of objects in the next two decades. 

Investigation of SBHBs has been spearheaded by theoretical studies which have investigated how black holes grow, form pairs and interact with their environment. They matured into a field that now faces questions rooted in thorough understanding of an extensive parameter space of SBHBs, their accretion flows and stellar environments. It became apparent that if further advances are to be made, observations of SBHBs must follow hand in hand, so to illuminate which portion of this parameter space is relevant for increasingly sophisticated theoretical models and simulations.

Observational searches for SBHB have however been challenging and while they unearthed a handful of binary candidates early on, their nature remained inconclusive in absence of the systematic multi-wavelength followup. They still provided valuable lessons as the present landscape of observational searches is represented by better designed, multi-wavelength and multi-year observational campaigns. The spectroscopic searches for SBHBs seem capable of delivering a statistically significant sample of binaries and their first results appear broadly consistent with theoretical predictions. In combination with other complementary observational techniques, they are well positioned to pave the way to a discovery of SBHBs. 
 
\begin{acknowledgement}
\begin{svgraybox}
T.B. acknowledges the support from the Alfred P. Sloan Foundation under Grant No. BR2013-016. This research was supported in part by the National Science Foundation under Grant No. NSF AST-1211677 and NSF AST-1333360.
\end{svgraybox}
\end{acknowledgement}

\bibliographystyle{spphys}
\bibliography{smbh}

\begin{thebibliography}{100}
\providecommand{\url}[1]{{#1}}
\providecommand{\urlprefix}{URL }
\expandafter\ifx\csname urlstyle\endcsname\relax
  \providecommand{\doi}[1]{DOI \discretionary{}{}{}#1}\else
  \providecommand{\doi}{DOI \discretionary{}{}{}\begingroup
  \urlstyle{rm}\Url}\fi

\bibitem{ferrarese00}
L.~{Ferrarese}, D.~{Merritt}, \apjl \textbf{539}, L9 (2000).
\newblock \doi{10.1086/312838}

\bibitem{gebhardt00}
K.~{Gebhardt}, R.~{Bender}, G.~{Bower}, A.~{Dressler}, S.M. {Faber}, A.V.
  {Filippenko}, R.~{Green}, C.~{Grillmair}, L.C. {Ho}, J.~{Kormendy}, T.R.
  {Lauer}, J.~{Magorrian}, J.~{Pinkney}, D.~{Richstone}, S.~{Tremaine}, \apjl
  \textbf{539}, L13 (2000).
\newblock \doi{10.1086/312840}

\bibitem{tremaine02}
S.~{Tremaine}, K.~{Gebhardt}, R.~{Bender}, G.~{Bower}, A.~{Dressler}, S.M.
  {Faber}, A.V. {Filippenko}, R.~{Green}, C.~{Grillmair}, L.C. {Ho},
  J.~{Kormendy}, T.R. {Lauer}, J.~{Magorrian}, J.~{Pinkney}, D.~{Richstone},
  \apj \textbf{574}, 740 (2002).
\newblock \doi{10.1086/341002}

\bibitem{white78}
S.D.M. {White}, \mnras \textbf{184}, 185 (1978)

\bibitem{schw86}
F.~{Schweizer}, Science \textbf{231}, 227 (1986).
\newblock \doi{10.1126/science.231.4735.227}

\bibitem{bh92}
J.E. {Barnes}, L.~{Hernquist}, \araa \textbf{30}, 705 (1992).
\newblock \doi{10.1146/annurev.aa.30.090192.003421}

\bibitem{bh96}
J.E. {Barnes}, L.~{Hernquist}, \apj \textbf{471}, 115 (1996).
\newblock \doi{10.1086/177957}

\bibitem{volonteri03}
M.~{Volonteri}, F.~{Haardt}, P.~{Madau}, \apj \textbf{582}, 559 (2003).
\newblock \doi{10.1086/344675}

\bibitem{dimatteo08a}
T.~{Di Matteo}, J.~{Colberg}, V.~{Springel}, L.~{Hernquist}, D.~{Sijacki}, \apj
  \textbf{676}, 33 (2008).
\newblock \doi{10.1086/524921}

\bibitem{bbr80}
M.C. {Begelman}, R.D. {Blandford}, M.J. {Rees}, \nat \textbf{287}, 307 (1980).
\newblock \doi{10.1038/287307a0}

\bibitem{mayer13}
L.~{Mayer}, Classical and Quantum Gravity \textbf{30}(24), 244008 (2013).
\newblock \doi{10.1088/0264-9381/30/24/244008}

\bibitem{chandra43}
S.~{Chandrasekhar}, \apj \textbf{97}, 255 (1943).
\newblock \doi{10.1086/144517}

\bibitem{ostriker99}
E.C. {Ostriker}, \apj \textbf{513}, 252 (1999).
\newblock \doi{10.1086/306858}

\bibitem{mm01}
M.~{Milosavljevi{\'c}}, D.~{Merritt}, \apj \textbf{563}, 34 (2001).
\newblock \doi{10.1086/323830}

\bibitem{escala04}
A.~{Escala}, R.B. {Larson}, P.S. {Coppi}, D.~{Mardones}, \apj \textbf{607}, 765
  (2004).
\newblock \doi{10.1086/386278}

\bibitem{fiacconi13}
D.~{Fiacconi}, L.~{Mayer}, R.~{Ro{\v s}kar}, M.~{Colpi}, \apjl \textbf{777},
  L14 (2013).
\newblock \doi{10.1088/2041-8205/777/1/L14}

\bibitem{escala05}
A.~{Escala}, R.B. {Larson}, P.S. {Coppi}, D.~{Mardones}, \apj \textbf{630}, 152
  (2005).
\newblock \doi{10.1086/431747}

\bibitem{dotti06}
M.~{Dotti}, M.~{Colpi}, F.~{Haardt}, \mnras \textbf{367}, 103 (2006).
\newblock \doi{10.1111/j.1365-2966.2005.09956.x}

\bibitem{dotti07}
M.~{Dotti}, M.~{Colpi}, F.~{Haardt}, L.~{Mayer}, \mnras \textbf{379}, 956
  (2007).
\newblock \doi{10.1111/j.1365-2966.2007.12010.x}

\bibitem{dotti09}
M.~{Dotti}, M.~{Ruszkowski}, L.~{Paredi}, M.~{Colpi}, M.~{Volonteri},
  F.~{Haardt}, \mnras \textbf{396}, 1640 (2009).
\newblock \doi{10.1111/j.1365-2966.2009.14840.x}

\bibitem{callegari11}
S.~{Callegari}, S.~{Kazantzidis}, L.~{Mayer}, M.~{Colpi}, J.M. {Bellovary},
  T.~{Quinn}, J.~{Wadsley}, \apj \textbf{729}, 85 (2011).
\newblock \doi{10.1088/0004-637X/729/2/85}

\bibitem{khan12}
F.M. {Khan}, I.~{Berentzen}, P.~{Berczik}, A.~{Just}, L.~{Mayer},
  K.~{Nitadori}, S.~{Callegari}, \apj \textbf{756}, 30 (2012).
\newblock \doi{10.1088/0004-637X/756/1/30}

\bibitem{chapon13}
D.~{Chapon}, L.~{Mayer}, R.~{Teyssier}, \mnras \textbf{429}, 3114 (2013).
\newblock \doi{10.1093/mnras/sts568}

\bibitem{callegari09}
S.~{Callegari}, L.~{Mayer}, S.~{Kazantzidis}, M.~{Colpi}, F.~{Governato},
  T.~{Quinn}, J.~{Wadsley}, \apjl \textbf{696}, L89 (2009).
\newblock \doi{10.1088/0004-637X/696/1/L89}

\bibitem{mayer07}
L.~{Mayer}, S.~{Kazantzidis}, P.~{Madau}, M.~{Colpi}, T.~{Quinn}, J.~{Wadsley},
  Science \textbf{316}, 1874 (2007).
\newblock \doi{10.1126/science.1141858}

\bibitem{berczik06}
P.~{Berczik}, D.~{Merritt}, R.~{Spurzem}, H.P. {Bischof}, \apjl \textbf{642},
  L21 (2006).
\newblock \doi{10.1086/504426}

\bibitem{preto11}
M.~{Preto}, I.~{Berentzen}, P.~{Berczik}, R.~{Spurzem}, \apjl \textbf{732}, L26
  (2011).
\newblock \doi{10.1088/2041-8205/732/2/L26}

\bibitem{khan11}
F.M. {Khan}, A.~{Just}, D.~{Merritt}, \apj \textbf{732}, 89 (2011).
\newblock \doi{10.1088/0004-637X/732/2/89}

\bibitem{khan12a}
F.M. {Khan}, M.~{Preto}, P.~{Berczik}, I.~{Berentzen}, A.~{Just}, R.~{Spurzem},
  \apj \textbf{749}, 147 (2012).
\newblock \doi{10.1088/0004-637X/749/2/147}

\bibitem{khan13}
F.M. {Khan}, K.~{Holley-Bockelmann}, P.~{Berczik}, A.~{Just}, \apj
  \textbf{773}, 100 (2013).
\newblock \doi{10.1088/0004-637X/773/2/100}

\bibitem{vasiliev14}
E.~{Vasiliev}, F.~{Antonini}, D.~{Merritt}, \apj \textbf{785}, 163 (2014).
\newblock \doi{10.1088/0004-637X/785/2/163}

\bibitem{an05}
P.J. {Armitage}, P.~{Natarajan}, \apj \textbf{634}, 921 (2005).
\newblock \doi{10.1086/497108}

\bibitem{macfadyen08}
A.I. {MacFadyen}, M.~{Milosavljevi{\'c}}, \apj \textbf{672}, 83 (2008).
\newblock \doi{10.1086/523869}

\bibitem{cuadra09}
J.~{Cuadra}, P.J. {Armitage}, R.D. {Alexander}, M.C. {Begelman}, \mnras
  \textbf{393}, 1423 (2009).
\newblock \doi{10.1111/j.1365-2966.2008.14147.x}

\bibitem{hayasaki09}
K.~{Hayasaki}, \pasj \textbf{61}, 65 (2009).
\newblock \doi{10.1093/pasj/61.1.65}

\bibitem{roedig12}
C.~{Roedig}, A.~{Sesana}, M.~{Dotti}, J.~{Cuadra}, P.~{Amaro-Seoane},
  F.~{Haardt}, \aap \textbf{545}, A127 (2012).
\newblock \doi{10.1051/0004-6361/201219986}

\bibitem{shi12}
J.M. {Shi}, J.H. {Krolik}, S.H. {Lubow}, J.F. {Hawley}, \apj \textbf{749}, 118
  (2012).
\newblock \doi{10.1088/0004-637X/749/2/118}

\bibitem{noble12}
S.C. {Noble}, B.C. {Mundim}, H.~{Nakano}, J.H. {Krolik}, M.~{Campanelli},
  Y.~{Zlochower}, N.~{Yunes}, \apj \textbf{755}, 51 (2012).
\newblock \doi{10.1088/0004-637X/755/1/51}

\bibitem{kocsis12a}
B.~{Kocsis}, Z.~{Haiman}, A.~{Loeb}, \mnras \textbf{427}, 2660 (2012).
\newblock \doi{10.1111/j.1365-2966.2012.22129.x}

\bibitem{kocsis12b}
B.~{Kocsis}, Z.~{Haiman}, A.~{Loeb}, \mnras \textbf{427}, 2680 (2012).
\newblock \doi{10.1111/j.1365-2966.2012.22118.x}

\bibitem{dorazio13}
D.J. {D'Orazio}, Z.~{Haiman}, A.~{MacFadyen}, \mnras \textbf{436}, 2997 (2013).
\newblock \doi{10.1093/mnras/stt1787}

\bibitem{farris14}
B.D. {Farris}, P.~{Duffell}, A.I. {MacFadyen}, Z.~{Haiman}, \apj \textbf{783},
  134 (2014).
\newblock \doi{10.1088/0004-637X/783/2/134}

\bibitem{lp79}
D.N.C. {Lin}, J.~{Papaloizou}, \mnras \textbf{186}, 799 (1979)

\bibitem{vanmeter10}
J.R. {van Meter}, J.H. {Wise}, M.C. {Miller}, C.S. {Reynolds}, J.~{Centrella},
  J.G. {Baker}, W.D. {Boggs}, B.J. {Kelly}, S.T. {McWilliams}, \apjl
  \textbf{711}, L89 (2010).
\newblock \doi{10.1088/2041-8205/711/2/L89}

\bibitem{palenzuela09}
C.~{Palenzuela}, M.~{Anderson}, L.~{Lehner}, S.L. {Liebling}, D.~{Neilsen},
  Physical Review Letters \textbf{103}(8), 081101 (2009).
\newblock \doi{10.1103/PhysRevLett.103.081101}

\bibitem{bode10}
T.~{Bode}, R.~{Haas}, T.~{Bogdanovi{\'c}}, P.~{Laguna}, D.~{Shoemaker}, \apj
  \textbf{715}, 1117 (2010).
\newblock \doi{10.1088/0004-637X/715/2/1117}

\bibitem{farris10}
B.D. {Farris}, Y.T. {Liu}, S.L. {Shapiro}, \prd \textbf{81}(8), 084008 (2010).
\newblock \doi{10.1103/PhysRevD.81.084008}

\bibitem{palenzuela10a}
C.~{Palenzuela}, L.~{Lehner}, S.~{Yoshida}, \prd \textbf{81}(8), 084007 (2010).
\newblock \doi{10.1103/PhysRevD.81.084007}

\bibitem{palenzuela10}
C.~{Palenzuela}, L.~{Lehner}, S.L. {Liebling}, Science \textbf{329}, 927
  (2010).
\newblock \doi{10.1126/science.1191766}

\bibitem{moesta10}
P.~{M{\"o}sta}, C.~{Palenzuela}, L.~{Rezzolla}, L.~{Lehner}, S.~{Yoshida},
  D.~{Pollney}, \prd \textbf{81}(6), 064017 (2010).
\newblock \doi{10.1103/PhysRevD.81.064017}

\bibitem{moesta12}
P.~{Moesta}, D.~{Alic}, L.~{Rezzolla}, O.~{Zanotti}, C.~{Palenzuela}, \apjl
  \textbf{749}, L32 (2012).
\newblock \doi{10.1088/2041-8205/749/2/L32}

\bibitem{bode12}
T.~{Bode}, T.~{Bogdanovi{\'c}}, R.~{Haas}, J.~{Healy}, P.~{Laguna},
  D.~{Shoemaker}, \apj \textbf{744}, 45 (2012).
\newblock \doi{10.1088/0004-637X/744/1/45}

\bibitem{farris11}
B.D. {Farris}, Y.T. {Liu}, S.L. {Shapiro}, \prd \textbf{84}(2), 024024 (2011).
\newblock \doi{10.1103/PhysRevD.84.024024}

\bibitem{alic12}
D.~{Alic}, P.~{Moesta}, L.~{Rezzolla}, O.~{Zanotti}, J.L. {Jaramillo}, \apj
  \textbf{754}, 36 (2012).
\newblock \doi{10.1088/0004-637X/754/1/36}

\bibitem{giaco12}
B.~{Giacomazzo}, J.G. {Baker}, M.C. {Miller}, C.S. {Reynolds}, J.R. {van
  Meter}, \apjl \textbf{752}, L15 (2012).
\newblock \doi{10.1088/2041-8205/752/1/L15}

\bibitem{gold14}
R.~{Gold}, V.~{Paschalidis}, Z.B. {Etienne}, S.L. {Shapiro}, H.P. {Pfeiffer},
  \prd \textbf{89}(6), 064060 (2014).
\newblock \doi{10.1103/PhysRevD.89.064060}

\bibitem{roedig11}
C.~{Roedig}, M.~{Dotti}, A.~{Sesana}, J.~{Cuadra}, M.~{Colpi}, \mnras
  \textbf{415}, 3033 (2011).
\newblock \doi{10.1111/j.1365-2966.2011.18927.x}

\bibitem{sesana11}
A.~{Sesana}, A.~{Gualandris}, M.~{Dotti}, \mnras \textbf{415}, L35 (2011).
\newblock \doi{10.1111/j.1745-3933.2011.01073.x}

\bibitem{key11}
J.S. {Key}, N.J. {Cornish}, \prd \textbf{83}(8), 083001 (2011).
\newblock \doi{10.1103/PhysRevD.83.083001}

\bibitem{as13}
P.~{Amaro-Seoane}, S.~{Aoudia}, S.~{Babak}, P.~{Bin{\'e}truy}, E.~{Berti},
  A.~{Boh{\'e}}, C.~{Caprini}, M.~{Colpi}, N.J. {Cornish}, K.~{Danzmann}, J.F.
  {Dufaux}, J.~{Gair}, I.~{Hinder}, O.~{Jennrich}, P.~{Jetzer}, A.~{Klein},
  R.N. {Lang}, A.~{Lobo}, T.~{Littenberg}, S.T. {McWilliams}, G.~{Nelemans},
  A.~{Petiteau}, E.K. {Porter}, B.F. {Schutz}, A.~{Sesana}, R.~{Stebbins},
  T.~{Sumner}, M.~{Vallisneri}, S.~{Vitale}, M.~{Volonteri}, H.~{Ward},
  B.~{Wardell}, GW Notes, Vol.~6, p.~4-110 \textbf{6}, 4 (2013)

\bibitem{reynolds13}
C.S. {Reynolds}, Classical and Quantum Gravity \textbf{30}(24), 244004 (2013).
\newblock \doi{10.1088/0264-9381/30/24/244004}

\bibitem{campanelli07}
M.~{Campanelli}, C.O. {Lousto}, Y.~{Zlochower}, D.~{Merritt}, Physical Review
  Letters \textbf{98}(23), 231102 (2007).
\newblock \doi{10.1103/PhysRevLett.98.231102}

\bibitem{lousto11}
C.O. {Lousto}, Y.~{Zlochower}, Physical Review Letters \textbf{107}(23), 231102
  (2011).
\newblock \doi{10.1103/PhysRevLett.107.231102}

\bibitem{bogdanovic07}
T.~{Bogdanovi{\'c}}, C.S. {Reynolds}, M.C. {Miller}, \apjl \textbf{661}, L147
  (2007).
\newblock \doi{10.1086/518769}

\bibitem{dotti10}
M.~{Dotti}, M.~{Volonteri}, A.~{Perego}, M.~{Colpi}, M.~{Ruszkowski},
  F.~{Haardt}, \mnras \textbf{402}, 682 (2010).
\newblock \doi{10.1111/j.1365-2966.2009.15922.x}

\bibitem{dotti13}
M.~{Dotti}, M.~{Colpi}, S.~{Pallini}, A.~{Perego}, M.~{Volonteri}, \apj
  \textbf{762}, 68 (2013).
\newblock \doi{10.1088/0004-637X/762/2/68}

\bibitem{sorathia13}
K.A. {Sorathia}, J.H. {Krolik}, J.F. {Hawley}, \apj \textbf{777}, 21 (2013).
\newblock \doi{10.1088/0004-637X/777/1/21}

\bibitem{miller13}
M.C. {Miller}, J.H. {Krolik}, \apj \textbf{774}, 43 (2013).
\newblock \doi{10.1088/0004-637X/774/1/43}

\bibitem{lodato13}
G.~{Lodato}, D.~{Gerosa}, \mnras \textbf{429}, L30 (2013).
\newblock \doi{10.1093/mnrasl/sls018}

\bibitem{fragile05}
P.C. {Fragile}, P.~{Anninos}, \apj \textbf{623}, 347 (2005).
\newblock \doi{10.1086/428433}

\bibitem{fragile07}
P.C. {Fragile}, O.M. {Blaes}, P.~{Anninos}, J.D. {Salmonson}, \apj
  \textbf{668}, 417 (2007).
\newblock \doi{10.1086/521092}

\bibitem{mckinney13}
J.C. {McKinney}, A.~{Tchekhovskoy}, R.D. {Blandford}, Science \textbf{339}, 49
  (2013).
\newblock \doi{10.1126/science.1230811}

\bibitem{bonning07}
E.W. {Bonning}, G.A. {Shields}, S.~{Salviander}, \apjl \textbf{666}, L13
  (2007).
\newblock \doi{10.1086/521674}

\bibitem{volonteri07}
M.~{Volonteri}, \apjl \textbf{663}, L5 (2007).
\newblock \doi{10.1086/519525}

\bibitem{schnittman07}
J.D. {Schnittman}, \apjl \textbf{667}, L133 (2007).
\newblock \doi{10.1086/522203}

\bibitem{sesana07}
A.~{Sesana}, \mnras \textbf{382}, L6 (2007).
\newblock \doi{10.1111/j.1745-3933.2007.00375.x}

\bibitem{loeb07}
A.~{Loeb}, Physical Review Letters \textbf{99}(4), 041103 (2007).
\newblock \doi{10.1103/PhysRevLett.99.041103}

\bibitem{gualandris08}
A.~{Gualandris}, D.~{Merritt}, \apj \textbf{678}, 780 (2008).
\newblock \doi{10.1086/586877}

\bibitem{komossa08a}
S.~{Komossa}, H.~{Zhou}, H.~{Lu}, \apjl \textbf{678}, L81 (2008).
\newblock \doi{10.1086/588656}

\bibitem{shields08}
G.A. {Shields}, E.W. {Bonning}, \apj \textbf{682}, 758 (2008).
\newblock \doi{10.1086/589427}

\bibitem{korn08}
D.A. {Kornreich}, R.V.E. {Lovelace}, \apj \textbf{681}, 104 (2008).
\newblock \doi{10.1086/587511}

\bibitem{berti08}
E.~{Berti}, M.~{Volonteri}, \apj \textbf{684}, 822 (2008).
\newblock \doi{10.1086/590379}

\bibitem{schnittman08}
J.D. {Schnittman}, J.H. {Krolik}, \apj \textbf{684}, 835 (2008).
\newblock \doi{10.1086/590363}

\bibitem{hb08}
K.~{Holley-Bockelmann}, K.~{G{\"u}ltekin}, D.~{Shoemaker}, N.~{Yunes}, \apj
  \textbf{686}, 829 (2008).
\newblock \doi{10.1086/591218}

\bibitem{komossa08}
S.~{Komossa}, D.~{Merritt}, \apjl \textbf{689}, L89 (2008).
\newblock \doi{10.1086/595883}

\bibitem{blecha08}
L.~{Blecha}, A.~{Loeb}, \mnras \textbf{390}, 1311 (2008).
\newblock \doi{10.1111/j.1365-2966.2008.13790.x}

\bibitem{volonteri08}
M.~{Volonteri}, P.~{Madau}, \apjl \textbf{687}, L57 (2008).
\newblock \doi{10.1086/593353}

\bibitem{oneill09}
S.M. {O'Neill}, M.C. {Miller}, T.~{Bogdanovi{\'c}}, C.S. {Reynolds}, J.D.
  {Schnittman}, \apj \textbf{700}, 859 (2009).
\newblock \doi{10.1088/0004-637X/700/1/859}

\bibitem{devecchi09}
B.~{Devecchi}, E.~{Rasia}, M.~{Dotti}, M.~{Volonteri}, M.~{Colpi}, \mnras
  \textbf{394}, 633 (2009).
\newblock \doi{10.1111/j.1365-2966.2008.14329.x}

\bibitem{merritt09}
D.~{Merritt}, J.D. {Schnittman}, S.~{Komossa}, \apj \textbf{699}, 1690 (2009).
\newblock \doi{10.1088/0004-637X/699/2/1690}

\bibitem{volonteri10}
M.~{Volonteri}, K.~{G{\"u}ltekin}, M.~{Dotti}, \mnras \textbf{404}, 2143
  (2010).
\newblock \doi{10.1111/j.1365-2966.2010.16431.x}

\bibitem{robinson10}
A.~{Robinson}, S.~{Young}, D.J. {Axon}, P.~{Kharb}, J.E. {Smith}, \apjl
  \textbf{717}, L122 (2010).
\newblock \doi{10.1088/2041-8205/717/2/L122}

\bibitem{lovelace10}
R.V.E. {Lovelace}, D.A. {Kornreich}, \mnras \textbf{402}, 2753 (2010).
\newblock \doi{10.1111/j.1365-2966.2009.16095.x}

\bibitem{corrales10}
L.R. {Corrales}, Z.~{Haiman}, A.~{MacFadyen}, \mnras \textbf{404}, 947 (2010).
\newblock \doi{10.1111/j.1365-2966.2010.16324.x}

\bibitem{guedes11}
J.~{Guedes}, P.~{Madau}, L.~{Mayer}, S.~{Callegari}, \apj \textbf{729}, 125
  (2011).
\newblock \doi{10.1088/0004-637X/729/2/125}

\bibitem{blecha11}
L.~{Blecha}, T.J. {Cox}, A.~{Loeb}, L.~{Hernquist}, \mnras \textbf{412}, 2154
  (2011).
\newblock \doi{10.1111/j.1365-2966.2010.18042.x}

\bibitem{nixon11}
C.J. {Nixon}, P.J. {Cossins}, A.R. {King}, J.E. {Pringle}, \mnras \textbf{412},
  1591 (2011).
\newblock \doi{10.1111/j.1365-2966.2010.17952.x}

\bibitem{nixon12}
C.J. {Nixon}, \mnras \textbf{423}, 2597 (2012).
\newblock \doi{10.1111/j.1365-2966.2012.21072.x}

\bibitem{lousto12}
C.O. {Lousto}, Y.~{Zlochower}, M.~{Dotti}, M.~{Volonteri}, \prd \textbf{85}(8),
  084015 (2012).
\newblock \doi{10.1103/PhysRevD.85.084015}

\bibitem{ponce12}
M.~{Ponce}, J.A. {Faber}, J.C. {Lombardi}, \apj \textbf{745}, 71 (2012).
\newblock \doi{10.1088/0004-637X/745/1/71}

\bibitem{komossa12}
S.~{Komossa}, Advances in Astronomy \textbf{2012}, 364973 (2012).
\newblock \doi{10.1155/2012/364973}

\bibitem{stone12}
N.~{Stone}, A.~{Loeb}, \mnras \textbf{422}, 1933 (2012).
\newblock \doi{10.1111/j.1365-2966.2012.20577.x}

\bibitem{shields13}
G.A. {Shields}, E.W. {Bonning}, \apjl \textbf{772}, L5 (2013).
\newblock \doi{10.1088/2041-8205/772/1/L5}

\bibitem{king13}
A.~{King}, C.~{Nixon}, Classical and Quantum Gravity \textbf{30}(24), 244006
  (2013).
\newblock \doi{10.1088/0264-9381/30/24/244006}

\bibitem{gerosa14}
D.~{Gerosa}, A.~{Sesana}, ArXiv e-prints  (2014)

\bibitem{volonteri09}
M.~{Volonteri}, J.M. {Miller}, M.~{Dotti}, \apjl \textbf{703}, L86 (2009).
\newblock \doi{10.1088/0004-637X/703/1/L86}

\bibitem{dotti12}
M.~{Dotti}, A.~{Sesana}, R.~{Decarli}, Advances in Astronomy \textbf{2012},
  940568 (2012).
\newblock \doi{10.1155/2012/940568}

\bibitem{schnittman13}
J.D. {Schnittman}, Classical and Quantum Gravity \textbf{30}(24), 244007
  (2013).
\newblock \doi{10.1088/0264-9381/30/24/244007}

\bibitem{rodriguez06}
C.~{Rodriguez}, G.B. {Taylor}, R.T. {Zavala}, A.B. {Peck}, L.K. {Pollack}, R.W.
  {Romani}, \apj \textbf{646}, 49 (2006).
\newblock \doi{10.1086/504825}

\bibitem{maness04}
H.L. {Maness}, G.B. {Taylor}, R.T. {Zavala}, A.B. {Peck}, L.K. {Pollack}, \apj
  \textbf{602}, 123 (2004).
\newblock \doi{10.1086/380919}

\bibitem{rodriguez09}
C.~{Rodriguez}, G.B. {Taylor}, R.T. {Zavala}, Y.M. {Pihlstr{\"o}m}, A.B.
  {Peck}, \apj \textbf{697}, 37 (2009).
\newblock \doi{10.1088/0004-637X/697/1/37}

\bibitem{morganti09}
R.~{Morganti}, B.~{Emonts}, T.~{Oosterloo}, \aap \textbf{496}, L9 (2009).
\newblock \doi{10.1051/0004-6361/200911705}

\bibitem{burke11}
S.~{Burke-Spolaor}, \mnras \textbf{410}, 2113 (2011).
\newblock \doi{10.1111/j.1365-2966.2010.17586.x}

\bibitem{condon11}
J.~{Condon}, J.~{Darling}, Y.Y. {Kovalev}, L.~{Petrov}, ArXiv e-prints  (2011)

\bibitem{burke14}
S.~{Burke-Spolaor}, A.~{Brazier}, S.~{Chatterjee}, J.~{Comerford}, J.~{Cordes},
  T.J.W. {Lazio}, X.~{Liu}, Y.~{Shen}, ArXiv e-prints  (2014)

\bibitem{fan98}
J.H. {Fan}, G.Z. {Xie}, E.~{Pecontal}, A.~{Pecontal}, Y.~{Copin}, \apj
  \textbf{507}, 173 (1998).
\newblock \doi{10.1086/306301}

\bibitem{rieger00}
F.M. {Rieger}, K.~{Mannheim}, \aap \textbf{359}, 948 (2000)

\bibitem{depaolis02}
F.~{De Paolis}, G.~{Ingrosso}, A.A. {Nucita}, \aap \textbf{388}, 470 (2002).
\newblock \doi{10.1051/0004-6361:20020519}

\bibitem{liu14}
F.K. {Liu}, S.~{Li}, S.~{Komossa}, \apj \textbf{786}, 103 (2014).
\newblock \doi{10.1088/0004-637X/786/2/103}

\bibitem{valtonen08}
M.J. {Valtonen}, H.J. {Lehto}, K.~{Nilsson}, J.~{Heidt}, L.O. {Takalo},
  A.~{Sillanp{\"a}{\"a}}, C.~{Villforth}, M.~{Kidger}, G.~{Poyner},
  T.~{Pursimo}, S.~{Zola}, J.H. {Wu}, X.~{Zhou}, K.~{Sadakane}, M.~{Drozdz},
  D.~{Koziel}, D.~{Marchev}, W.~{Ogloza}, C.~{Porowski}, M.~{Siwak},
  G.~{Stachowski}, M.~{Winiarski}, V.P. {Hentunen}, M.~{Nissinen}, A.~{Liakos},
  S.~{Dogru}, \nat \textbf{452}, 851 (2008).
\newblock \doi{10.1038/nature06896}

\bibitem{bogdanovic08}
T.~{Bogdanovi{\'c}}, B.D. {Smith}, S.~{Sigurdsson}, M.~{Eracleous}, \apjs
  \textbf{174}, 455 (2008).
\newblock \doi{10.1086/521828}

\bibitem{al96}
P.~{Artymowicz}, S.H. {Lubow}, \apjl \textbf{467}, L77 (1996).
\newblock \doi{10.1086/310200}

\bibitem{gr00}
A.~{Gould}, H.W. {Rix}, \apjl \textbf{532}, L29 (2000).
\newblock \doi{10.1086/312562}

\bibitem{hayasaki07}
K.~{Hayasaki}, S.~{Mineshige}, H.~{Sudou}, \pasj \textbf{59}, 427 (2007).
\newblock \doi{10.1093/pasj/59.2.427}

\bibitem{bogdanovic09a}
T.~{Bogdanovi{\'c}}, M.~{Eracleous}, S.~{Sigurdsson}, \apj \textbf{697}, 288
  (2009).
\newblock \doi{10.1088/0004-637X/697/1/288}

\bibitem{dotti09a}
M.~{Dotti}, C.~{Montuori}, R.~{Decarli}, M.~{Volonteri}, M.~{Colpi},
  F.~{Haardt}, \mnras \textbf{398}, L73 (2009).
\newblock \doi{10.1111/j.1745-3933.2009.00714.x}

\bibitem{bl09}
T.A. {Boroson}, T.R. {Lauer}, \nat \textbf{458}, 53 (2009).
\newblock \doi{10.1038/nature07779}

\bibitem{tang09}
S.~{Tang}, J.~{Grindlay}, \apj \textbf{704}, 1189 (2009).
\newblock \doi{10.1088/0004-637X/704/2/1189}

\bibitem{decarli10}
R.~{Decarli}, M.~{Dotti}, C.~{Montuori}, T.~{Liimets}, A.~{Ederoclite}, \apjl
  \textbf{720}, L93 (2010).
\newblock \doi{10.1088/2041-8205/720/1/L93}

\bibitem{barrows11}
R.S. {Barrows}, C.H.S. {Lacy}, D.~{Kennefick}, J.~{Kennefick}, M.S. {Seigar},
  \na \textbf{16}, 122 (2011).
\newblock \doi{10.1016/j.newast.2010.08.004}

\bibitem{tsal11}
P.~{Tsalmantza}, R.~{Decarli}, M.~{Dotti}, D.W. {Hogg}, \apj \textbf{738}, 20
  (2011).
\newblock \doi{10.1088/0004-637X/738/1/20}

\bibitem{tsai13}
C.W. {Tsai}, T.H. {Jarrett}, D.~{Stern}, B.~{Emonts}, R.S. {Barrows}, R.J.
  {Assef}, R.P. {Norris}, P.R.M. {Eisenhardt}, C.~{Lonsdale}, A.W. {Blain},
  D.J. {Benford}, J.~{Wu}, B.~{Stalder}, C.W. {Stubbs}, F.W. {High}, K.L. {Li},
  A.K.H. {Kong}, \apj \textbf{779}, 41 (2013).
\newblock \doi{10.1088/0004-637X/779/1/41}

\bibitem{shields09}
G.A. {Shields}, D.J. {Rosario}, K.L. {Smith}, E.W. {Bonning}, S.~{Salviander},
  J.S. {Kalirai}, R.~{Strickler}, E.~{Ramirez-Ruiz}, A.A. {Dutton}, T.~{Treu},
  P.J. {Marshall}, \apj \textbf{707}, 936 (2009).
\newblock \doi{10.1088/0004-637X/707/2/936}

\bibitem{civano10}
F.~{Civano}, M.~{Elvis}, G.~{Lanzuisi}, K.~{Jahnke}, G.~{Zamorani},
  L.~{Blecha}, A.~{Bongiorno}, M.~{Brusa}, A.~{Comastri}, H.~{Hao},
  A.~{Leauthaud}, A.~{Loeb}, V.~{Mainieri}, E.~{Piconcelli}, M.~{Salvato},
  N.~{Scoville}, J.~{Trump}, C.~{Vignali}, T.~{Aldcroft}, M.~{Bolzonella},
  E.~{Bressert}, A.~{Finoguenov}, A.~{Fruscione}, A.M. {Koekemoer},
  N.~{Cappelluti}, F.~{Fiore}, S.~{Giodini}, R.~{Gilli}, C.D. {Impey}, S.J.
  {Lilly}, E.~{Lusso}, S.~{Puccetti}, J.D. {Silverman}, H.~{Aussel},
  P.~{Capak}, D.~{Frayer}, E.~{Le Floch}, H.J. {McCracken}, D.B. {Sanders},
  D.~{Schiminovich}, Y.~{Taniguchi}, \apj \textbf{717}, 209 (2010).
\newblock \doi{10.1088/0004-637X/717/1/209}

\bibitem{lusso14}
E.~{Lusso}, R.~{Decarli}, M.~{Dotti}, C.~{Montuori}, D.W. {Hogg},
  P.~{Tsalmantza}, M.~{Fumagalli}, J.X. {Prochaska}, \mnras \textbf{441}, 316
  (2014).
\newblock \doi{10.1093/mnras/stu572}

\bibitem{bogdanovic09}
T.~{Bogdanovi{\'c}}, M.~{Eracleous}, S.~{Sigurdsson}, \nar \textbf{53}, 113
  (2009).
\newblock \doi{10.1016/j.newar.2009.09.005}

\bibitem{popovic12}
L.{\v C}. {Popovi{\'c}}, \nar \textbf{56}, 74 (2012).
\newblock \doi{10.1016/j.newar.2011.11.001}

\bibitem{eracleous12}
M.~{Eracleous}, T.A. {Boroson}, J.P. {Halpern}, J.~{Liu}, \apjs \textbf{201},
  23 (2012).
\newblock \doi{10.1088/0067-0049/201/2/23}

\bibitem{bon12}
E.~{Bon}, P.~{Jovanovi{\'c}}, P.~{Marziani}, A.I. {Shapovalova}, N.~{Bon},
  V.~{Borka Jovanovi{\'c}}, D.~{Borka}, J.~{Sulentic}, L.{\v C}. {Popovi{\'c}},
  \apj \textbf{759}, 118 (2012).
\newblock \doi{10.1088/0004-637X/759/2/118}

\bibitem{decarli13}
R.~{Decarli}, M.~{Dotti}, M.~{Fumagalli}, P.~{Tsalmantza}, C.~{Montuori},
  E.~{Lusso}, D.W. {Hogg}, J.X. {Prochaska}, \mnras \textbf{433}, 1492 (2013).
\newblock \doi{10.1093/mnras/stt831}

\bibitem{liu13}
X.~{Liu}, Y.~{Shen}, F.~{Bian}, A.~{Loeb}, S.~{Tremaine}, ArXiv e-prints
  (2013)

\bibitem{shen13}
Y.~{Shen}, X.~{Liu}, A.~{Loeb}, S.~{Tremaine}, \apj \textbf{775}, 49 (2013).
\newblock \doi{10.1088/0004-637X/775/1/49}

\bibitem{ju13}
W.~{Ju}, J.E. {Greene}, R.R. {Rafikov}, S.J. {Bickerton}, C.~{Badenes}, \apj
  \textbf{777}, 44 (2013).
\newblock \doi{10.1088/0004-637X/777/1/44}

\bibitem{mathes14}
G.~{Mathes}, M.~{Eracleous}, S.~{Sigurdsson}, J.C. {Runnoe}, T.~{Bogdanovic},
  in \emph{American Astronomical Society Meeting Abstracts}, \emph{American
  Astronomical Society Meeting Abstracts}, vol. 223 (2014), \emph{American
  Astronomical Society Meeting Abstracts}, vol. 223, p. 250.15

\bibitem{bentz08}
M.C. {Bentz}, J.L. {Walsh}, A.J. {Barth}, N.~{Baliber}, N.~{Bennert},
  G.~{Canalizo}, A.V. {Filippenko}, M.~{Ganeshalingam}, E.L. {Gates}, J.E.
  {Greene}, M.G. {Hidas}, K.D. {Hiner}, N.~{Lee}, W.~{Li}, M.A. {Malkan},
  T.~{Minezaki}, F.J.D. {Serduke}, J.H. {Shiode}, J.M. {Silverman}, T.N.
  {Steele}, D.~{Stern}, R.A. {Street}, C.E. {Thornton}, T.~{Treu}, X.~{Wang},
  J.H. {Woo}, Y.~{Yoshii}, \apjl \textbf{689}, L21 (2008).
\newblock \doi{10.1086/595719}

\bibitem{bentz09}
M.C. {Bentz}, J.L. {Walsh}, A.J. {Barth}, N.~{Baliber}, V.N. {Bennert},
  G.~{Canalizo}, A.V. {Filippenko}, M.~{Ganeshalingam}, E.L. {Gates}, J.E.
  {Greene}, M.G. {Hidas}, K.D. {Hiner}, N.~{Lee}, W.~{Li}, M.A. {Malkan},
  T.~{Minezaki}, Y.~{Sakata}, F.J.D. {Serduke}, J.M. {Silverman}, T.N.
  {Steele}, D.~{Stern}, R.A. {Street}, C.E. {Thornton}, T.~{Treu}, X.~{Wang},
  J.H. {Woo}, Y.~{Yoshii}, \apj \textbf{705}, 199 (2009).
\newblock \doi{10.1088/0004-637X/705/1/199}

\bibitem{bentz10}
M.C. {Bentz}, J.L. {Walsh}, A.J. {Barth}, Y.~{Yoshii}, J.H. {Woo}, X.~{Wang},
  T.~{Treu}, C.E. {Thornton}, R.A. {Street}, T.N. {Steele}, J.M. {Silverman},
  F.J.D. {Serduke}, Y.~{Sakata}, T.~{Minezaki}, M.A. {Malkan}, W.~{Li},
  N.~{Lee}, K.D. {Hiner}, M.G. {Hidas}, J.E. {Greene}, E.L. {Gates},
  M.~{Ganeshalingam}, A.V. {Filippenko}, G.~{Canalizo}, V.N. {Bennert},
  N.~{Baliber}, \apj \textbf{716}, 993 (2010).
\newblock \doi{10.1088/0004-637X/716/2/993}

\bibitem{bentz10a}
M.C. {Bentz}, K.~{Horne}, A.J. {Barth}, V.N. {Bennert}, G.~{Canalizo}, A.V.
  {Filippenko}, E.L. {Gates}, M.A. {Malkan}, T.~{Minezaki}, T.~{Treu}, J.H.
  {Woo}, J.L. {Walsh}, \apjl \textbf{720}, L46 (2010).
\newblock \doi{10.1088/2041-8205/720/1/L46}

\bibitem{pancoast11}
A.~{Pancoast}, B.J. {Brewer}, T.~{Treu}, \apj \textbf{730}, 139 (2011).
\newblock \doi{10.1088/0004-637X/730/2/139}

\bibitem{barth11}
A.J. {Barth}, A.~{Pancoast}, S.J. {Thorman}, V.N. {Bennert}, D.J. {Sand},
  W.~{Li}, G.~{Canalizo}, A.V. {Filippenko}, E.L. {Gates}, J.E. {Greene}, M.A.
  {Malkan}, D.~{Stern}, T.~{Treu}, J.H. {Woo}, R.J. {Assef}, H.J. {Bae}, B.J.
  {Brewer}, T.~{Buehler}, S.B. {Cenko}, K.I. {Clubb}, M.C. {Cooper}, A.M.
  {Diamond-Stanic}, K.D. {Hiner}, S.F. {H{\"o}nig}, M.D. {Joner}, M.T.
  {Kandrashoff}, C.D. {Laney}, M.S. {Lazarova}, A.M. {Nierenberg}, D.~{Park},
  J.M. {Silverman}, D.~{Son}, A.~{Sonnenfeld}, E.J. {Tollerud}, J.L. {Walsh},
  R.~{Walters}, R.L. {da Silva}, M.~{Fumagalli}, M.D. {Gregg}, C.E. {Harris},
  E.Y. {Hsiao}, J.~{Lee}, L.~{Lopez}, J.~{Rex}, N.~{Suzuki}, J.R. {Trump},
  D.~{Tytler}, G.~{Worseck}, H.M. {Yesuf}, \apjl \textbf{743}, L4 (2011).
\newblock \doi{10.1088/2041-8205/743/1/L4}

\bibitem{dietrich12}
M.~{Dietrich}, B.M. {Peterson}, C.J. {Grier}, M.C. {Bentz}, J.~{Eastman},
  S.~{Frank}, R.~{Gonzalez}, J.L. {Marshall}, D.L. {DePoy}, J.L. {Prieto}, \apj
  \textbf{757}, 53 (2012).
\newblock \doi{10.1088/0004-637X/757/1/53}

\bibitem{pancoast13}
A.~{Pancoast}, B.J. {Brewer}, T.~{Treu}, D.~{Park}, A.J. {Barth}, M.C. {Bentz},
  J.H. {Woo}, ArXiv e-prints  (2013)

\bibitem{zoghbi12}
A.~{Zoghbi}, A.C. {Fabian}, C.S. {Reynolds}, E.M. {Cackett}, \mnras
  \textbf{422}, 129 (2012).
\newblock \doi{10.1111/j.1365-2966.2012.20587.x}

\bibitem{kara13a}
E.~{Kara}, A.C. {Fabian}, E.M. {Cackett}, J.F. {Steiner}, P.~{Uttley}, D.R.
  {Wilkins}, A.~{Zoghbi}, \mnras \textbf{428}, 2795 (2013).
\newblock \doi{10.1093/mnras/sts155}

\bibitem{kara13b}
E.~{Kara}, A.C. {Fabian}, E.M. {Cackett}, G.~{Miniutti}, P.~{Uttley}, \mnras
  \textbf{430}, 1408 (2013).
\newblock \doi{10.1093/mnras/stt024}

\bibitem{kara14}
E.~{Kara}, E.M. {Cackett}, A.C. {Fabian}, C.~{Reynolds}, P.~{Uttley}, \mnras
  \textbf{439}, L26 (2014).
\newblock \doi{10.1093/mnrasl/slt173}

\bibitem{cackett14}
E.M. {Cackett}, A.~{Zoghbi}, C.~{Reynolds}, A.C. {Fabian}, E.~{Kara},
  P.~{Uttley}, D.R. {Wilkins}, \mnras \textbf{438}, 2980 (2014).
\newblock \doi{10.1093/mnras/stt2424}

\bibitem{uttley14}
P.~{Uttley}, E.M. {Cackett}, A.C. {Fabian}, E.~{Kara}, D.R. {Wilkins}, ArXiv
  e-prints  (2014)

\bibitem{fabian89}
A.C. {Fabian}, M.J. {Rees}, L.~{Stella}, N.E. {White}, \mnras \textbf{238}, 729
  (1989)

\bibitem{sesana12}
A.~{Sesana}, C.~{Roedig}, M.T. {Reynolds}, M.~{Dotti}, \mnras \textbf{420}, 860
  (2012).
\newblock \doi{10.1111/j.1365-2966.2011.20097.x}

\bibitem{jovanovic13}
P.~{Jovanovi{\'c}}, V.~{Borka Jovanovi{\'c}}, D.~{Borka}, T.~{Bogdanovi{\'c}},
  ArXiv e-prints  (2013)

\bibitem{mckernan13}
B.~{McKernan}, K.E.S. {Ford}, B.~{Kocsis}, Z.~{Haiman}, \mnras \textbf{432},
  1468 (2013).
\newblock \doi{10.1093/mnras/stt567}

\bibitem{sesana08}
A.~{Sesana}, A.~{Vecchio}, C.N. {Colacino}, \mnras \textbf{390}, 192 (2008).
\newblock \doi{10.1111/j.1365-2966.2008.13682.x}

\bibitem{shannon13}
R.M. {Shannon}, V.~{Ravi}, W.A. {Coles}, G.~{Hobbs}, M.J. {Keith}, R.N.
  {Manchester}, J.S.B. {Wyithe}, M.~{Bailes}, N.D.R. {Bhat},
  S.~{Burke-Spolaor}, J.~{Khoo}, Y.~{Levin}, S.~{Oslowski}, J.M. {Sarkissian},
  W.~{van Straten}, J.P.W. {Verbiest}, J.B. {Want}, Science \textbf{342}, 334
  (2013)

\bibitem{sesana13}
A.~{Sesana}, \mnras \textbf{433}, L1 (2013).
\newblock \doi{10.1093/mnrasl/slt034}

\bibitem{burke13}
S.~{Burke-Spolaor}, Classical and Quantum Gravity \textbf{30}(22), 224013
  (2013).
\newblock \doi{10.1088/0264-9381/30/22/224013}

\bibitem{arzu14}
Z.~{Arzoumanian}, A.~{Brazier}, S.~{Burke-Spolaor}, S.J. {Chamberlin},
  S.~{Chatterjee}, J.M. {Cordes}, P.B. {Demorest}, X.~{Deng}, T.~{Dolch}, J.A.
  {Ellis}, R.D. {Ferdman}, N.~{Garver-Daniels}, F.~{Jenet}, G.~{Jones}, V.M.
  {Kaspi}, M.~{Koop}, M.~{Lam}, T.J.W. {Lazio}, A.N. {Lommen}, D.R. {Lorimer},
  J.~{Luo}, R.S. {Lynch}, D.R. {Madison}, M.~{McLaughlin}, S.T. {McWilliams},
  D.J. {Nice}, N.~{Palliyaguru}, T.T. {Pennucci}, S.M. {Ransom}, A.~{Sesana},
  X.~{Siemens}, I.H. {Stairs}, D.R. {Stinebring}, K.~{Stovall}, J.~{Swiggum},
  M.~{Vallisneri}, R.~{van Haasteren}, Y.~{Wang}, W.W. {Zhu}, ArXiv e-prints
  (2014)

\bibitem{babak12}
S.~{Babak}, A.~{Sesana}, \prd \textbf{85}(4), 044034 (2012).
\newblock \doi{10.1103/PhysRevD.85.044034}

\bibitem{ravi12}
V.~{Ravi}, J.S.B. {Wyithe}, G.~{Hobbs}, R.M. {Shannon}, R.N. {Manchester},
  D.R.B. {Yardley}, M.J. {Keith}, \apj \textbf{761}, 84 (2012).
\newblock \doi{10.1088/0004-637X/761/2/84}

\bibitem{tanaka13}
T.L. {Tanaka}, Z.~{Haiman}, Classical and Quantum Gravity \textbf{30}(22),
  224012 (2013).
\newblock \doi{10.1088/0264-9381/30/22/224012}

\bibitem{petiteau13}
A.~{Petiteau}, S.~{Babak}, A.~{Sesana}, M.~{de Ara{\'u}jo}, \prd
  \textbf{87}(6), 064036 (2013).
\newblock \doi{10.1103/PhysRevD.87.064036}

\end{thebibliography}

\end{document}